\documentclass[mathleft
]{an}
\usepackage{graphicx}
\usepackage{times}
\begin{document}

\Pagespan{789}{}
\Yearpublication{2006}%
\Yearsubmission{2005}%
\Month{11}%
\Volume{999}%
\Issue{88}%

\title{Photometry and spectroscopy of the newly discovered \\
eclipsing binary GSC 4589-2999}

\author{A.Liakos\inst{1}\fnmsep\thanks{Corresponding author:
  \email{alliakos@phys.uoa.gr}\newline}
\and  P.Bonfini\inst{2,3} \and  P.Niarchos\inst{1} \and
D.Hatzidimitriou\inst{1,3}}
\titlerunning{Photometry and spectroscopy of the EB GSC 4589:2999}
\authorrunning{Liakos, A., Bonfini, P., Niarchos, P. and Hatzidimitriou, D.}
\institute{Department of Astrophysics, Astronomy and Mechanics,
University of Athens, GR 157 84, Zografos, Athens, Hellas \and
Department of Physics, University of Crete, PO Box 2208, 71003,
Heraklion, Hellas \and IESL, Foundation for Research and
Technology-Hellas, 71110, Heraklion, Hellas}

\received{30 May 2005}
\accepted{11 Nov 2005}
\publonline{later}

\keywords{stars: binaries: eclipsing -- stars: fundamental
parameters -- stars: individual (GSC 4589-2999) -- techniques:
photometric -- techniques: spectroscopic}

\abstract{%
New CCD light curves of the recently detected eclipsing variable
GSC 4589-2999 were obtained and analysed using the Wilson-Deninney
code. Spectroscopic observations of the system allowed the
spectral classification of the components and the determination of
their radial velocities. The physical properties and absolute
parameters of the components and an updated ephemeris of the
system are given.}

\maketitle

\section{Introduction and Observations}

Eclipsing binary systems offer unique conditions to measure
fundamental parameters of stars, such as stellar masses, radii and
luminosities, which are of great importance in stu-\\dies of
stellar structure and evolution.

Using extensive photometric and spectroscopic data, we performed a
comprehensive study of the newly discovered eclipsing system GSC
4589-2999 (=TYC 4589-2999-1,\\
$\alpha_{2000}=20^{h}15^{m}0.2^{s},~\delta_{2000}
=+76^{\circ}54'18.2''$, V=10.6 mag). The light variability report
(as a by-product of observations of the eclipsing system EG Cep),
the first light curves and ephemeris of the system were published
by Liakos \& Niarchos (2007).

The new photometric observations of the system were carried out at
the Gerostathopoulion Observatory of the Un-\\iversity of Athens,
during 44 nights in June-October 2007 and 12 nights in June-July
2008, using the 0.4m Cassegrain telescope equipped with the
ST-8XMEI CCD camera and the BVRI Bessell photometric filters.

The spectroscopic observations were obtained with the 1.3m
Ritchey-Cretien telescope equipped with a 2000$\times$800 ISA SITe
CCD camera at Skinakas Observatory, located on the island of
Crete, Hellas, on September 2007, September 2008 and May 2009. For
the spectral classification, we used the Low Resolution
Spectrograph which is incorporated in the Focal Reducer
instrument. For the spectral classification we used a reflection
grating of 1302 lines/mm giving a nominal dispersion of 1.04
\AA/pixel, and an appropriate wavelength coverage (see Table 2) to
include both H$_{\alpha}$ and H$_{\beta}$ Balmer lines. For the
radial velocity (hereafter RVs) measurements a 2400 lines/mm
grating was used, giving a nominal dispersion of 0.55 \AA/pixel,
while the wavelength range was selected to include the Mg I
triplet and other metallic lines.

\section{Data reduction}

\subsection{Photometry}

Differential magnitudes were obtained using the software
\emph{MuniWin} v.1.1.23 (Hroch 1998). The magnitudes have not been
transformed to a standard system and the colour extinction was not
taken into account since standard star observations were not
performed. Information about the comparison stars can be found in
Table 1.

\begin{table}

\caption{The photometric comparison and check stars. The
magnitudes information are given in The Tycho-2 Catalogue (H$\o$g
2000)} \label{tab1}  \centering {

 \begin{tabular}{cccccc}
 \hline
 Star            &      B$_{TYC}$    &     V$_{TYC}$   & \emph{remarks}   \\
 \hline
 GSC 4589-2999   &       11.33 (6)   &     10.61 (4) &     variable     \\
 GSC 4589-2984   &       11.86 (8)   &     10.93 (6) &    comparison    \\
 GSC 4589-2842   &       13.4 (3)    &     11.6 (1)  &      check       \\
 \hline
 \end{tabular}}
 \end{table}

\subsection{Spectroscopy}

Various exposure times for the observations of 2007, 2008, 2009
were used (for details see Table 2). Before and after each
on-target observation, an arc calibration exposure (Ne- HeAr) was
recorded. Arc spectra were extracted from the arc exposures by
applying exactly the same profiles as to the corresponding object
spectra and they were used to calibrate the object spectra.

For the spectral classification procedure, a total of 44 standards
of spectral types ranging from O8 to M6 were observed with an
identical instrument configuration as the targets: 15 were
observed on September 29, 2007, 14 were observed on 31 August 2008
and another 15 were taken from Hatzidimitriou et al. (2006). All
standard star spectra were reduced and calibrated in exactly the
same way as the targets, following the procedure described in the
following. The data reduction was performed using the \emph{IRAF}
package 2.13-BETA2 (2006). The frames were bias subtracted and the
sky background was removed. The spectra were subsequently traced
and extracted using the all-in-one subpackage \emph{apextract}.

For the determination of the RVs, the softwares
\emph{\textbf{Ra}dial \textbf{Ve}locity \textbf{re}ductions}
v.2.1d (Nelson 2009) and \emph{Broadening Functions} v.2.4c
(Nelson 2009) were used.

\begin{table}

\caption{The spectroscopic observations log}

\label{tab2}

\centering{

 \begin{tabular}{ccccc}
   \hline

   Date of          &       UT  &   Phase    &  Exposure    &Wavelength        \\
 observation        &           &            & time (sec)   & coverage (\AA)   \\
   \hline
   29-9-2007        &   17:36   &   0.544    &   300        & 4728-6828        \\
   29-9-2007        &   17:42   &   0.547    &   300        & 4728-6828        \\
   29-9-2007        &   19:28   &   0.590    &   300        & 4728-6828        \\
   29-9-2007        &   19:35   &   0.593    &   300        & 4728-6828        \\
   3-9-2008         &   01:44   &   0.497    &   600        & 4713-6792        \\
   12-5-2009        &   23:54   &   0.688    &   1800       & 4535-5621        \\
   13-5-2009        &   21:19   &   0.216    &   1800       & 4535-5621        \\
   14-5-2009        &   20:09   &   0.780    &   1800       & 4535-5621        \\
   14-5-2009        &   20:49   &   0.795    &   1800       & 4535-5621        \\
   \hline

 \end{tabular}}
 \end{table}

\section{Data analysis}

\subsection{Spectroscopic analysis}

On September 2007 the spectra were obtained on the rising branch
of the secondary minimum, and particularly the second set of
spectra includes the spectra of both stars. On September 2008 the
spectroscopic observations were obtained during the total eclipse,
thus obtaining a clear spectrum for the primary (larger, hotter
and more massive) component. The exact phase moments when the
spectra were taken are given in Table 2. The spectrum of the
secondary component was obtained by the subtraction of the two
spectra (the spectra of 2008 from those of 2007), after
normalizing them to the same continuum, line-free point. The
resulting spectrum is of quite high signal-to-noise and clearly
indicates a cooler star (see Fig. 1b), as it will be discussed
below.

The classification of the spectra was achieved in two steps:

(i) Cross-correlation with the 44 standard star spectra, using the
method described in detail in Hatzidimitriou et al. (2006). The
accuracy of the spectral classification achieved with this method
depends on the fineness of the grid of standard spectra used and
on the signal-to-noise ratio of the cross-correlated spectra, and
it is found to be around 0.3-0.4 of a spectral type (cf. Bonfini
et al. 2009).

(ii) Visual detailed comparison of spectral features with standard
spectra around the spectral type indicated by step (i). This
procedure required prior normalization of the continua of the
spectra, and leads to a confirmation and fine-tuning of the
cross-correlation result.

Using steps (i) and (ii), we estimate that the primary component
of the system, namely GSC 4589-2999a (spectrum obtained in
September 2008), is of spectral type G1.5\\ $\pm$0.5. The spectrum
is shown in Fig. 1a (solid black line), along with the spectrum of
the G2 standard star HD 196755 and of the G0 standard SF-18, for
comparison. The spectrum of the secondary component, GSC
4589-2999b, resulted from the subtraction of the spectra taken in
2007 and 2008, is of K4 type, with an uncertainty of one subclass.
The spectrum is shown in Fig. 1b together with the comparison
spectra of the K4 star HD 198550 and the K5 star 32 Vul. The cross
correlation diagrams for each component are overplotted in Fig.
1c.

For the RVs determination, the Broadening Functions (hereafter
BFs) method (Rucinski 2002) on the spectra of 2009 was used. We
cropped all spectra in order to avoid the broad H$_{\beta}$ line,
and we included all the sharp metallic lines between 4865-5355
\AA. Each RV value and its error was derived statistically (mean
value and error) from the respective velocities resulted from BFs
method by using different standard stars. The standard stars were
HIP 40497, HIP 61317, HIP 65721 which are of F7, G0 and G5
spectral type, respectively. The heliocentric RVs with their
errors in parentheses are given in Table 3, and the RVs plot is
illustrated in Fig. 5.

\begin{table}

\caption{The heliocentric radial velocities measurements}

\label{tab3}

\centering {
 \begin{tabular}{ccccc}

\hline
HJD - 2450000       &   Phase    &          V$_1$       &       V$_2$           \\
                    &            &        (Km/sec)      &      (Km/sec)         \\
\hline
4964.5048           &   0.688    &         51 (29)      &      -165 (17)        \\
4965.3971           &   0.216    &        -93 (24)      &       156 (29)        \\
4966.3490           &   0.780    &         57 (24)      &      -180 (18)        \\
4966.3747           &   0.795    &         71 (32)      &      -170 (16)        \\

\hline

 \end{tabular}}
 \end{table}

\begin{figure}[t]
\centering{
\begin{tabular}{cr}
\includegraphics[width=7cm]{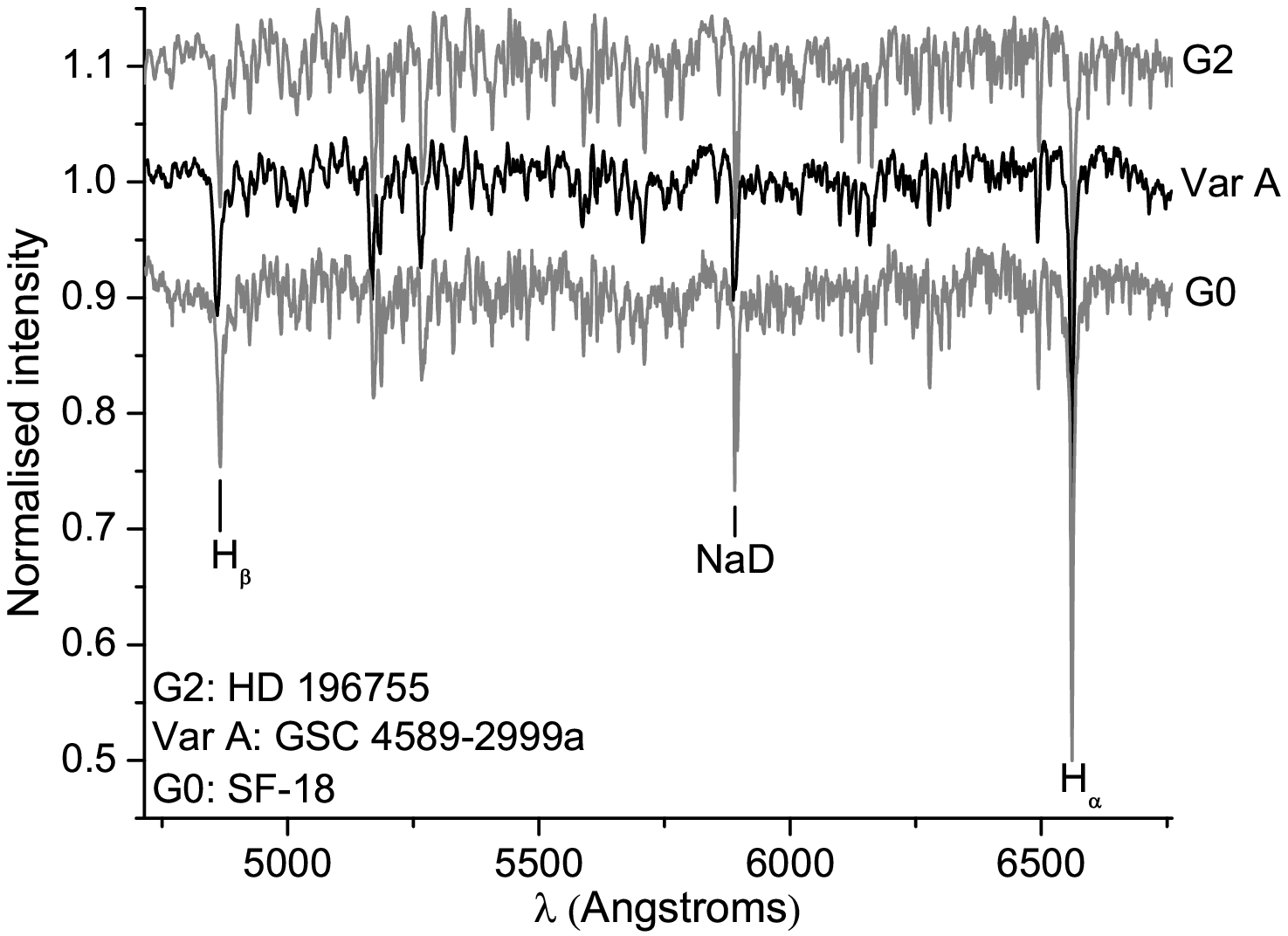}&(a)\\
\includegraphics[width=7cm]{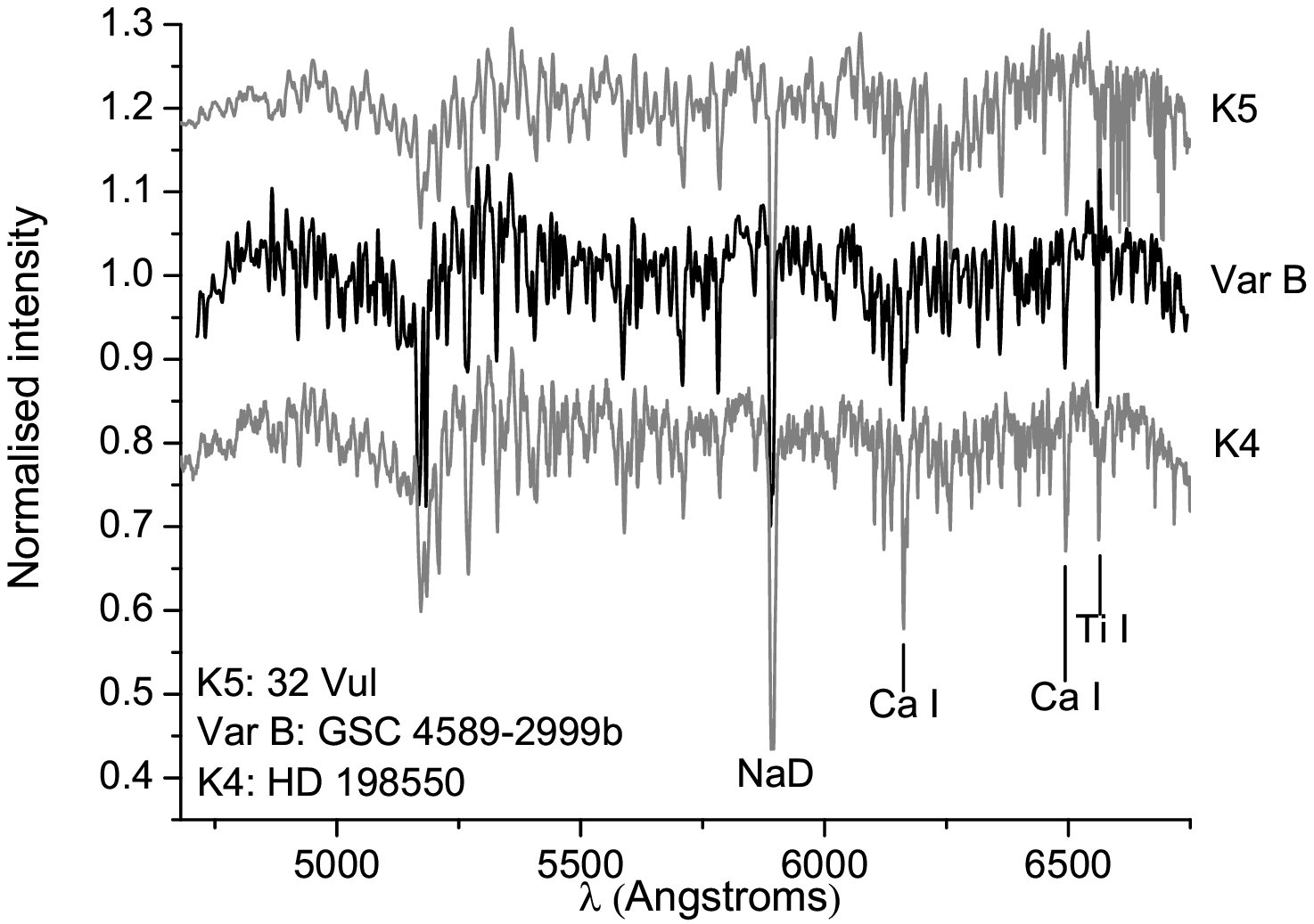}&(b)\\
\includegraphics[width=7cm]{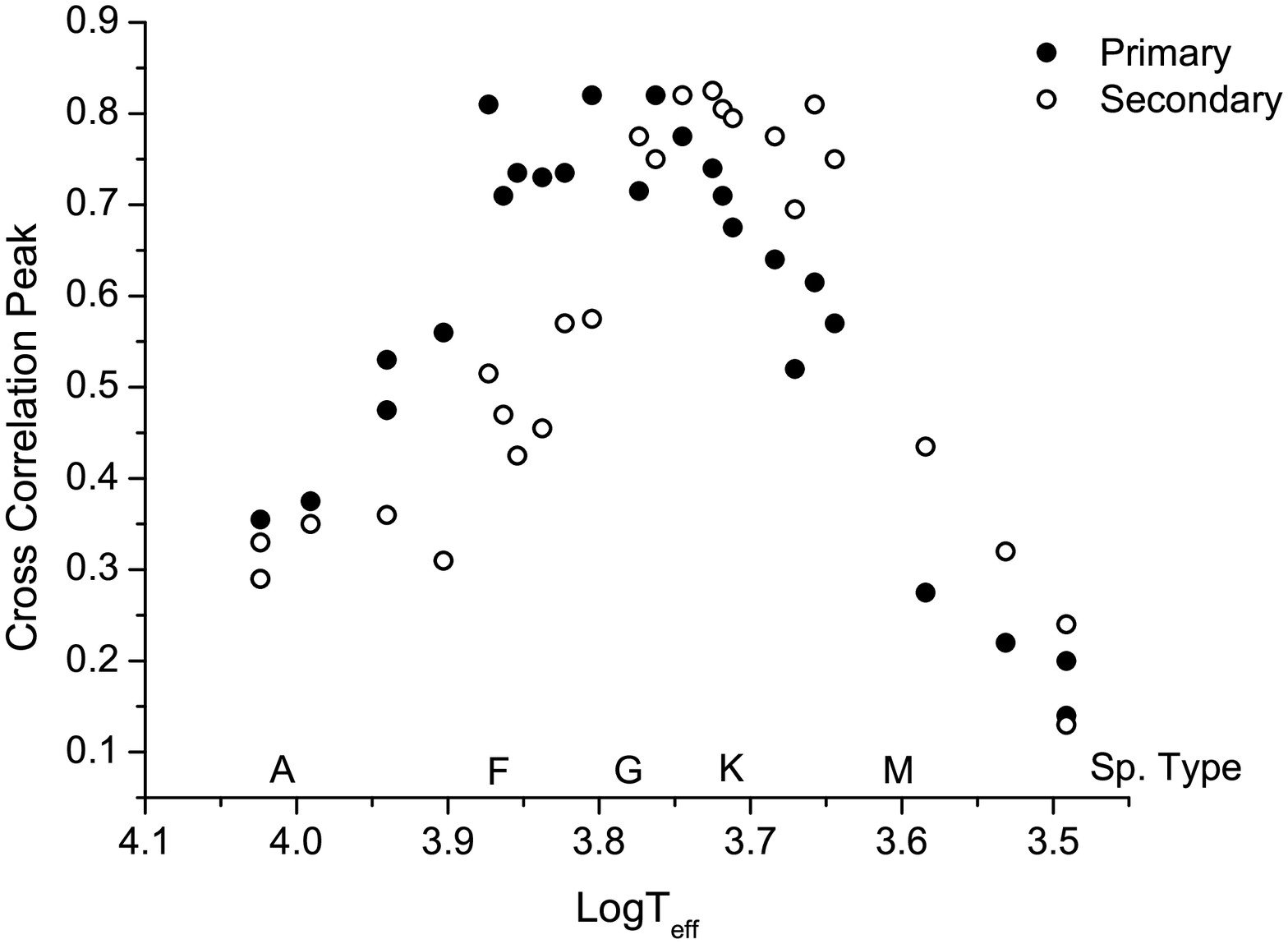}&(c)
\label{fig1}
\end{tabular}
\caption{The spectra of the primary (a) and secondary (b)
components of GSC 4589-2999. Information about the standard stars
are given in the text. (c) The Cross Correlation plots for each
component of the system.}}
\end{figure}

\subsection{Photometric and radial velocity analysis}

The updated ephemeris of the system was calculated by applying the
least squares method on all the available minima timings (Liakos
\& Niarchos 2009) and it is given by the following relation:

\textbf{Min.I=HJD 2454642.4992(2)+1.688653(1)$^{d}\times$~E}

The light curves and the RVs curves were analyzed with the
\emph{PHOEBE} v.0.29d software (Pr\v{s}a \& Zwitter 2005),
whi-\\ch uses the 2003 version of the Wilson-Devinney code (Wilson
\& Devinney 1971; Wilson 1979, 1990). The code was applied in MODE
2 (detached system) and all light (BVRI) and RV curves were
analysed simultaneously.

The temperature of the primary $T_1$ was fixed to the typical value of a
G1.5 star, namely 5830 K, while the temperature of the secondary $T_2$
was given initially the value 4550 K, a typical value for a K4
star, but during the iterations it was left as free parameter. Given the spectroscopic error of the primary's spectral type (see section 3.1), its corresponding temperature has an accuracy of 40 K.  Hence, the error of the $T_2$ value was calculated according to the error propagation method.
The albedo of each component, namely $A_1$ and $A_2$, and the gravity
darkening coefficients, $g_1$ and $g_2$, were assigned theoretical
values, according to the spectral types of the components. The
values of limb darkening coefficients, $x_1$ and $x_2$, were taken
from the tables of van Hamme (1993). The dimensionless potentials
$\Omega_{1}$ and $\Omega_{2}$, the mass ratio $q$, the fractional
luminosity $L_{1}$ of the primary component, the systemic radial
velocity $V_0$, the semi major axis \emph{a} and the inclination
of the system's orbit $i$ were treated as adjustable parameters.
Due to asymmetries near the primary maxima of the light curves, a
cool spot on the surface of the cooler component was assumed and
its parameters (latitude $Co-lat$, longitude $Co-lon$, radius $R$
and temperature factor $T_{spot}/T_{sur}$) were also adjusted.
Another analysis without the cool spot was also performed, but the
result was worse. The 'spotted' solution resulted in a sum of
squared residuals $\sum x^2$=0.0032, while the 'unspotted' one
yielded to a value of 0.0059, therefore we chose to adopt the
'spotted' solution. The comparison between the models including or
not the spot is presented in Fig. 3. The maximum radial velocity
of each component, namely $K_1$ and $K_2$, was calculated by
fitting a sinusoidal function on the points of each RV curve. The
synthetic and observed light and RV curves are shown in Figs 2 and
5, respectively, the 3D representation is illustrated in Fig. 4,
and the derived parameters are listed in Table 4.

\begin{figure}

\centering{

\begin{tabular}{c}
\includegraphics[width=7.5cm]{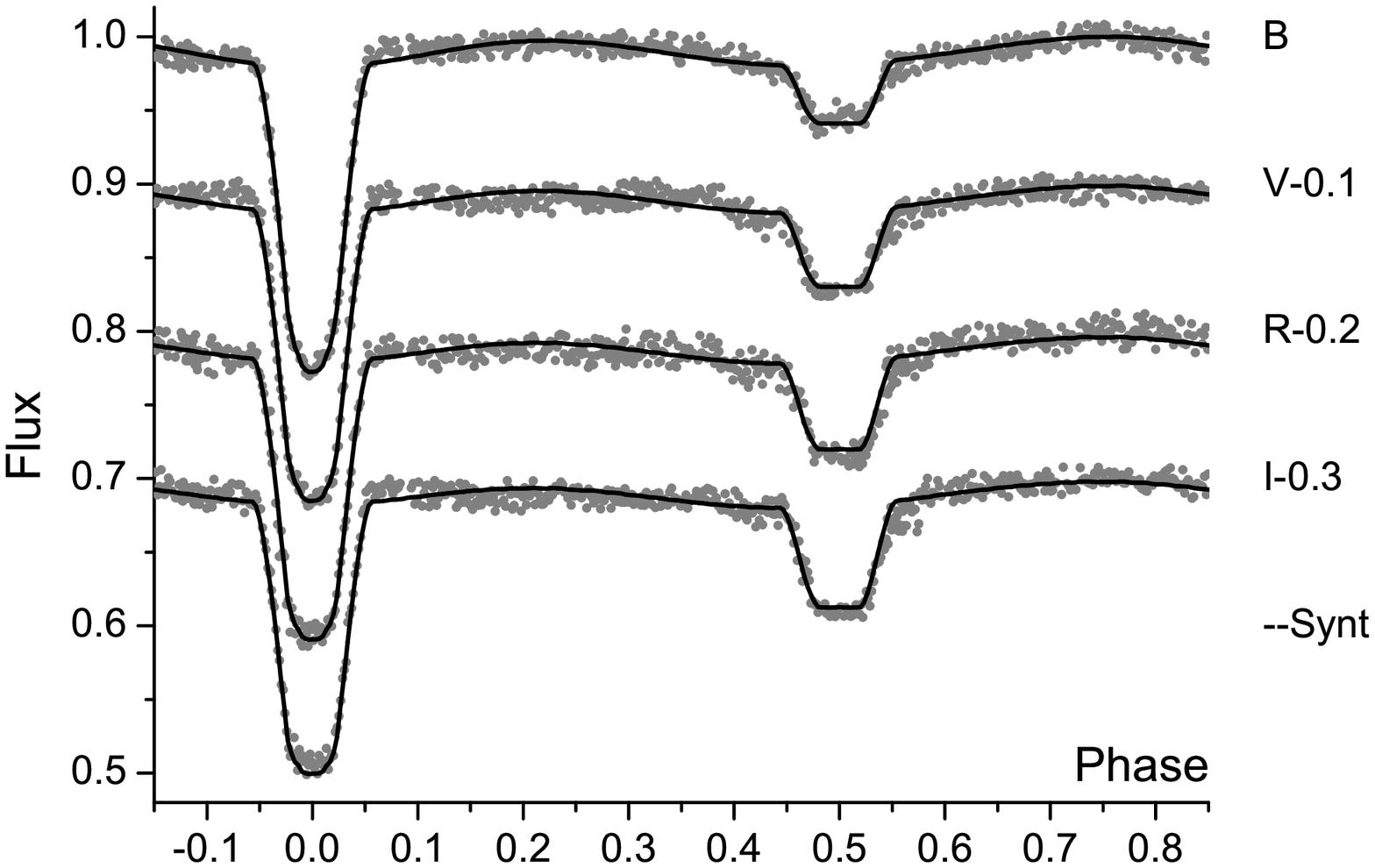}\\
\end{tabular}
\label{fig2} \caption{Synthetic (solid lines) and observed
(points) light curves of GSC 4589-2999.}

\begin{tabular}{c}
\includegraphics[width=7.5cm]{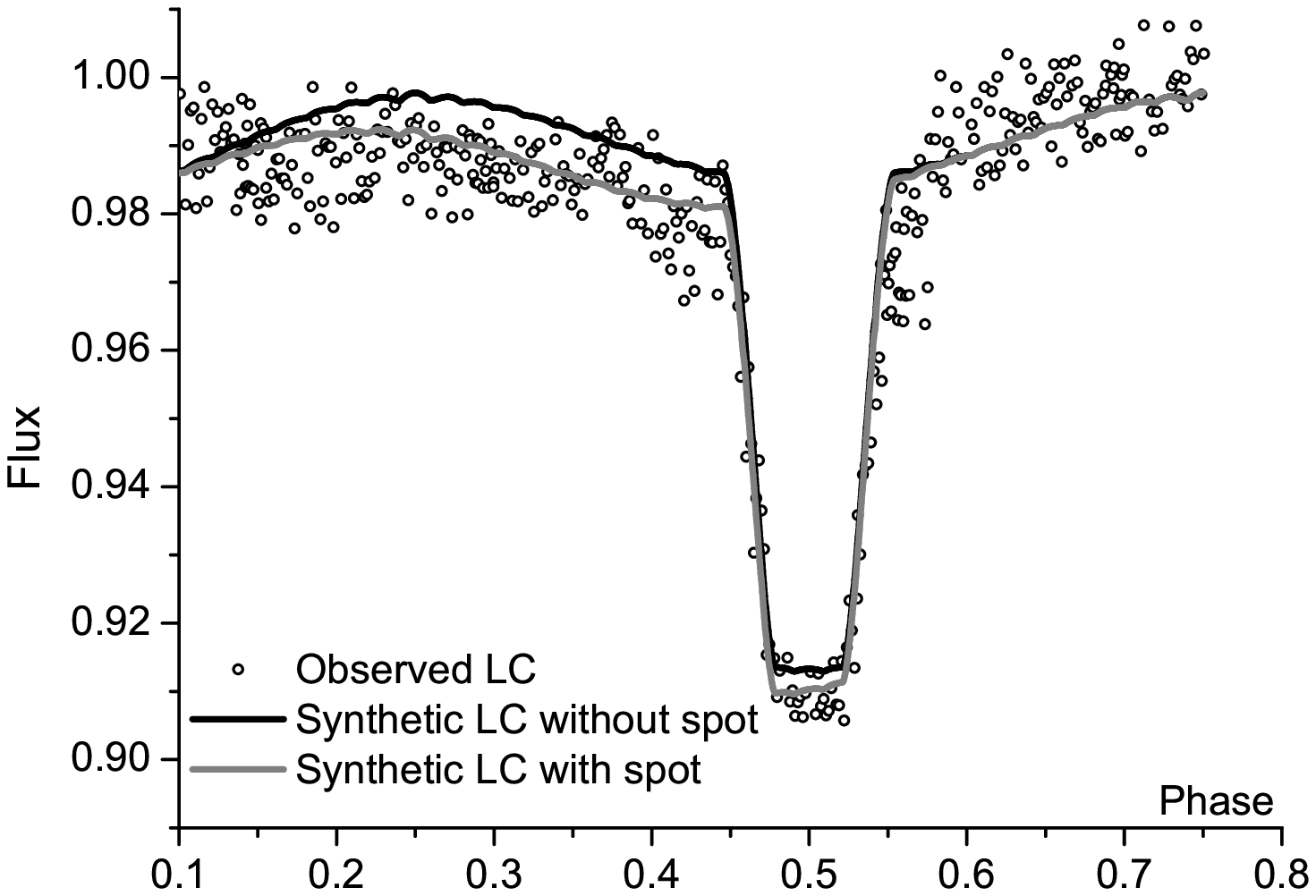}\\
\end{tabular}
\label{fig3} \caption{The comparison between the fit on the data
points of the synthetic light curves in I filter for spotted
(grey solid line) and unspotted (black solid line) models.}

\begin{tabular}{cc}
\includegraphics[width=3.2cm]{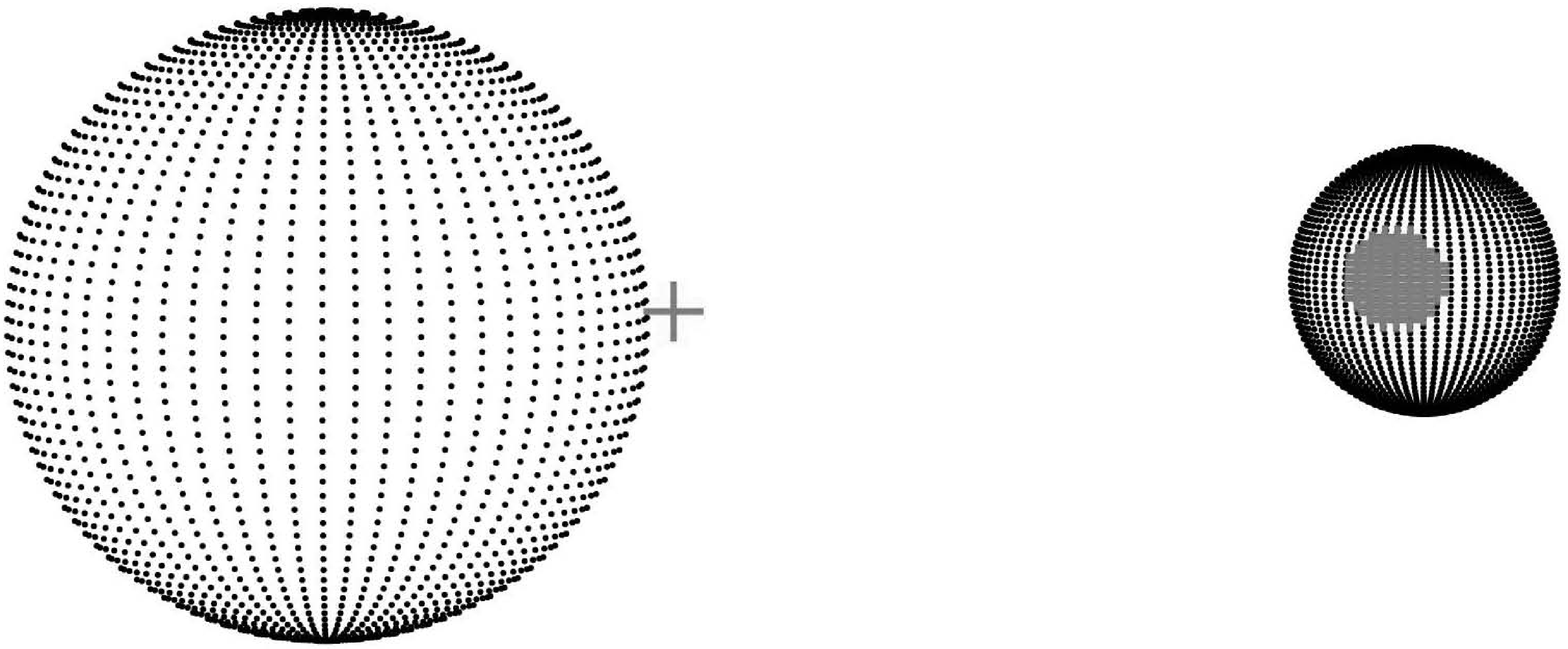}&\includegraphics[width=4.2cm]{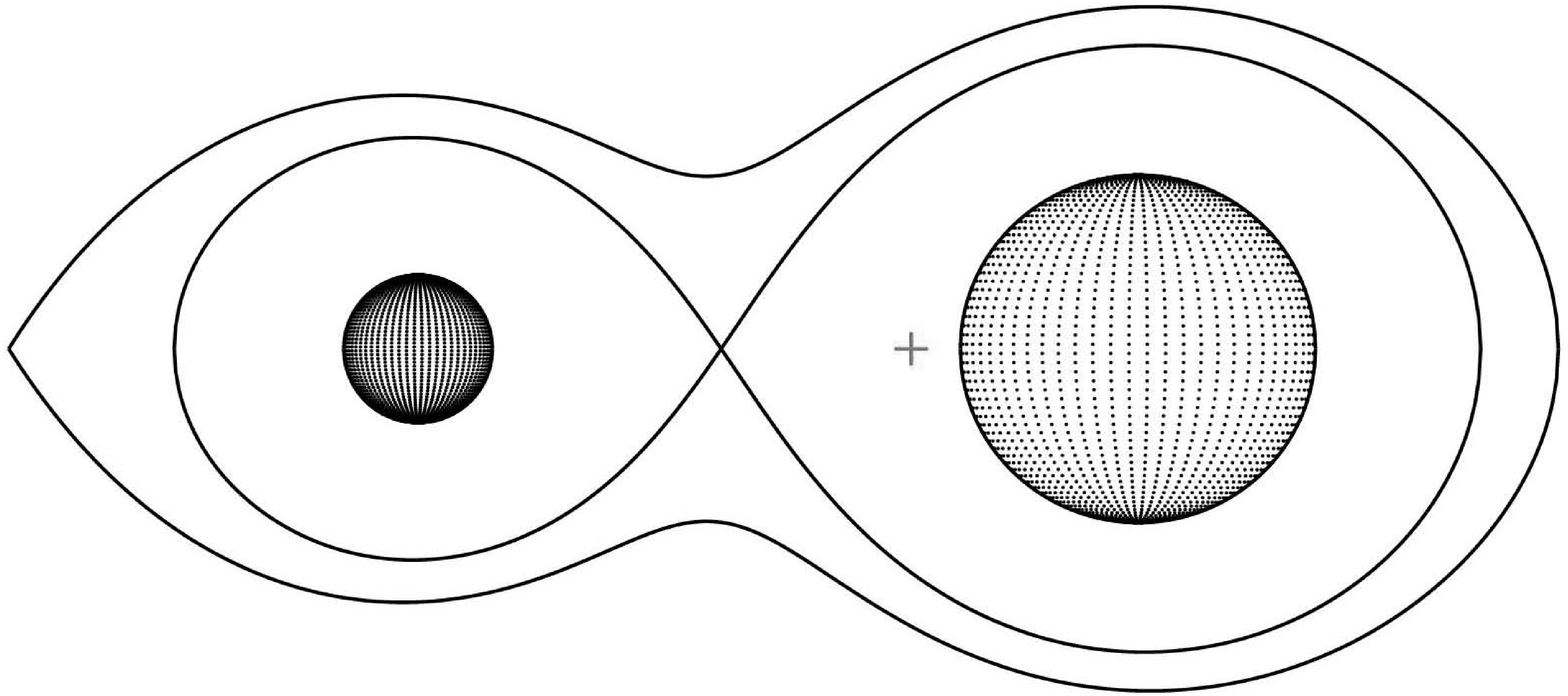}\\
                        (a)               &                 (b)                      \\
\end{tabular}
\label{fig4}\caption{The 3D model of GSC 4589-2999 at the phases:
(a) 0.34 where the cool spot is indicated and (b) 0.75 where the
center of mass (+) and inner and outer Roche Lobes are shown.}}

\end{figure}

\begin{figure}
\centering{
\includegraphics[width=7cm]{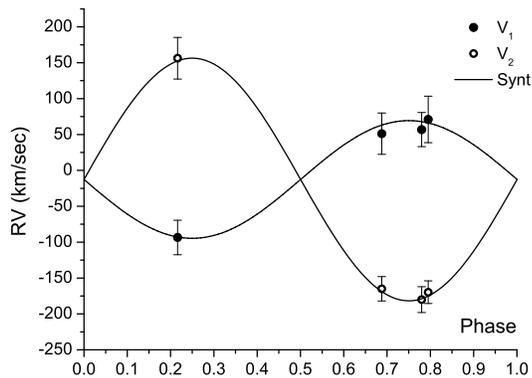}
\label{fig5} \caption{Synthetic (solid lines) and observed
(points) radial velocity curves of GSC 4589-2999.}}
\end{figure}

\begin{table}

\caption{The parameters (par.) of GSC 4589-2999 derived from the
light and radial velocities curves solution. P denotes the primary
and S the secondary component, respectively. The errors are given
in parentheses and correspond to the last digit of each value.}
\label{tab4}
\centering
\begin{tabular}{lc lc}
\hline
        Parameter               &       Value   &       Parameter                   &    Value  \\
\hline
i (deg)                     	&	86.3 (2)	&	x$_{1,B}$                   	&	0.731	\\
q ($m_{2}/m_{1}$)           	&	0.46 (3)	&	x$_{2,B}$                   	&	0.898	\\
$a$ (R$_\odot$)             	&	8.3 (3)	    &	x$_{1,V}$                   	&	0.597	\\
K$_1$ (km/sec)              	&	78 (3)	    &	x$_{2,V}$                   	&	0.758	\\
K$_2$ (km/sec)              	&	169 (1)	    &	x$_{1,R}$                   	&	0.514	\\
V$_0$ (km/sec)              	&	-13 (3)	    &	x$_{2,R}$                   	&	0.658	\\
T$_1$ (K)                   	&	5830 (40)	&	x$_{1,I}$                   	&	0.434	\\
T$_2$ (K)                   	&	4616 (116)	&	x$_{2,I}$                   	&	0.549	\\
\cline{3-4}
$\Omega_{1}$                	&	4.59 (1)	&    \multicolumn{2}{c}{\emph{Spot}} 		    \\
\cline{3-4}
$\Omega_{2}$                	&	5.75 (2)	&	Co-lat (deg)                 	&	86 (6)	\\
L$_1$/(L$_1$+L$_2$)$_B$     	&	0.961 (1)	&	Co-lon (deg)                	&	44 (2)	\\
L$_1$/(L$_1$+L$_2$)$_V$     	&	0.949 (1)	&	R (deg)                     	&	20 (6)	\\
L$_1$/(L$_1$+L$_2$)$_R$     	&	0.940 (1)	&	T$_{spot}$/T$_{sur}$        	&	0.6 (2)	\\
\cline{3-4}
L$_1$/(L$_1$+L$_2$)$_I$     	&	0.929 (1)	&   \multicolumn{2}{c}{\emph{Fractional radii}} \\
\cline{3-4}		
L$_2$/(L$_1$+L$_2$)$_B$     	&	0.039 (1)	&	r$_1$ (pole)                	&	0.24	\\
L$_2$/(L$_1$+L$_2$)$_V$     	&	0.051 (1)	&	r$_1$ (point)               	&	0.25	\\
L$_2$/(L$_1$+L$_2$)$_R$     	&	0.060 (1)	&	r$_1$ (side)                	&	0.24	\\
L$_2$/(L$_1$+L$_2$)$_I$     	&	0.071 (1)	&	r$_1$ (back)                	&	0.25	\\
g$_1$         	                &	0.32	    &	r$_2$ (pole)                	&	0.10    \\
g$_2$         	                &	0.32	    &	r$_2$ (point)               	&	0.10    \\
A$_1$  	                        &	0.5       	&   r$_2$ (side)                	&	0.10    \\
A$_2$  	                        &	0.5	        &	r$_2$ (back)                	&	0.10    \\
\hline
\end{tabular}
\end{table}

\section{Absolute elements of the components}

The geometric and photometric elements, derived from the
simultaneous analysis of the light and RV curves, were used to
compute the absolute elements of the components. These elements,
listed in Table 5, are used to place the two components on the
Mass-Radius (M--R) diagram (Fig. 6) in order to examine their
evolutionary status.

Both components lie inside the Terminal Age Main Sequence (TAMS)
and Zero Age Main Sequence (ZAMS) limits, revealing their
Main-Sequence (MS) nature. However, the primary component is
closer to TAMS line, while the secondary one is almost in the
middle of the limits. This result is in agreement with the theory
of stellar evolution according to their initial mass, since the
more massive component is found to be more evolved than the
secondary, assuming a simultaneous birth.

\begin{table}

\caption{The absolute elements of the components (P: Primary, S:
Secondary). The errors are given in parentheses.}
\label{tab5}

\centering{

\begin{tabular}{lccccc}
\hline
            &       M       &        R      &   T$_{eff}$ &         L      & M$_{bol}$      \\
            &  (M$_\odot$)  &  (R$_\odot$)  &      (K)    &  (L$_\odot$)   &  (mag)         \\
\hline
       P    &   1.8 (1)     &    2.02 (4)   &  5830 (40)  &    4.2 (2)     &  3.2 (3)       \\
       S    &   0.8 (1)     &    0.85 (2)   &  4616 (116) &    0.3 (1)     &  6.1 (8)       \\
\hline
\end{tabular}}
\end{table}

\begin{figure}
\centering{
\includegraphics[width=7cm]{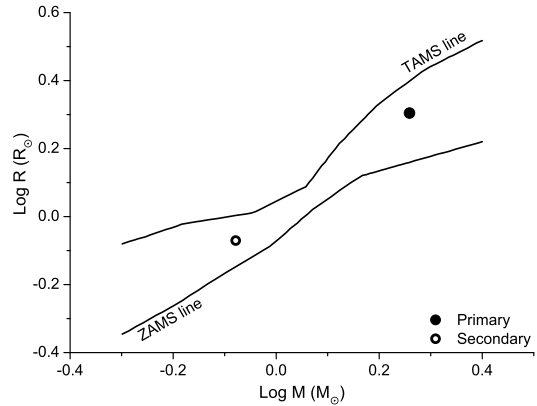}
\label{fig6} \caption{The position of the components of GSC
4589-2999 in the M--R diagram.}}
\end{figure}

\section{Discussion and conclusions}

The main contribution of the present paper is the first
determination of the physical parameters of the newly discovered
eclipsing binary system GSC 4589-2999. This study is based on
high-quality spectroscopic and multicolour CCD observations.

The two components of the system are classified as G1.5\\ $\pm$0.5
(primary) and K4$\pm$1 (secondary), respectively. The light curve
analysis yielded a temperature of the secondary component as 4616
K, in very good agreement, within the error limit, with the
temperature corresponding to the spectral type (K3/5, 4685 K -
4415 K), derived from our spectroscopic observations.

The shape of the light curves indicates an Algol-type eclipsing
binary with a period close to 1.69 d. A simultaneous analysis of
the light and radial velocity curves resulted in a detached
configuration with zero eccentricity. A small asymmetry near the
primary maximum was better explained by assuming a cool spot on
the surface of the secondary component, but it is well known that
spot solution suffers from non-uniqueness.

The derived absolute elements were used to study the evolutionary
status of the two components. It was found that both components
belong to the MS, while the primary component (more massive and
hotter) is more evolved than the secondary.

\acknowledgements
This work has been financially supported  by the
Special Account for Research Grants No 70/4/9709 of the National
\& Kapodistrian University of Athens, Hellas. We thank P. Reig for
the spectroscopic observations of September of 2008. Part of the
work presented here was done while P. Bonfini was still a student
at the University degli Studi di Milano - Bicocca, Italy and was
possible thanks to the Erasmus Program. Skinakas Observatory is a
collaborative project of the University of Crete, and the
Foundation for Research and Technology-Hellas.


\end{document}